\documentclass[twocolumn,groupedaddress,floatfix,prl]{revtex4}
\usepackage{bm}
\usepackage{graphicx}

\begin{document}

\title{Peptide Folding Kinetics from Replica Exchange Molecular Dynamics}

\author{Nicolae-Viorel Buchete}
%\email{buchete@alum.bu.edu}
%\altaffiliation{Current address: School of Physics, University College Dublin, Belfield, Dublin 4, Ireland; E-mail: buchete@alum.bu.edu}
\altaffiliation{Current address: School of Physics, University College Dublin, Dublin 4, Ireland; E-mail: buchete@alum.bu.edu}
\affiliation{Laboratory of Chemical Physics, National Institute of Diabetes and Digestive and Kidney Diseases, National Institutes of Health, Building 5, Bethesda, MD 20892-0520, USA}
\author{Gerhard Hummer}
\email{gerhard.hummer@nih.gov}
\affiliation{Laboratory of Chemical Physics, National Institute of Diabetes and Digestive and Kidney Diseases, National Institutes of Health, Building 5, Bethesda, MD 20892-0520, USA}

\begin{abstract}
We show how accurate kinetic information, such as the rates of protein folding and unfolding, can be extracted from replica-exchange molecular dynamics (REMD) simulations.  From the brief and continuous trajectory segments between replica exchanges, we estimate short-time propagators in conformation space and use them to construct a master equation.  For a helical peptide in explicit water, we determine the rates of transitions both locally between microscopic conformational states and globally for folding and unfolding. We show that accurate rates in the $\sim$1/(100~ns) to $\sim$1/(1~ns) range can be obtained from REMD with exchange times of 5 ps, in excellent agreement with results from long equilibrium molecular dynamics.
\end{abstract}

\pacs{}
 
\maketitle

Replica exchange molecular dynamics (REMD) \cite{Okamoto:CPL:1999:remd,GarciaSanbonmatsu:PNAS} is a powerful method to enhance the conformational sampling, addressing a serious challenge in molecular simulations \cite{BerneStraub:COSB:1997}.  Multiple non-interacting copies (or ``replicas'') of the system are simulated in parallel, each at a different temperature. To transfer the barrier-crossing efficiency from runs at high temperature to those at low temperature, configuration exchanges are attempted periodically (e.g., at time intervals $\delta t_\mathrm{REMD}$) between replicas at different temperatures ($T_i$ and $T_j$).  Those exchange attempts are accepted with a Metropolis probability  $P_\mathrm{REMD}(i \leftrightarrow j) =  \min \{ 1, \exp [ (\beta_j - \beta_i)(U_j-U_i) ] \}$ that enforces detailed balance and maintains canonical distributions at each temperature [with $U_i$ the potential energy of the $i$-th replica, $\beta_i =  1/(k_B T_i)$, and $k_B$ the Boltzmann constant].  After an accepted exchange, the particle velocities are appropriately re-scaled to the new temperature, or re-drawn from respective Maxwell-Boltzmann distributions.  Through a series of exchanges, high-temperature conformations are transferred occasionally to low temperature runs, facilitating the exploration of new configuration-space regions.

While enhancing the exploration of conformation space, REMD apparently does not permit the extraction of useful kinetic information.  Conformation exchanges result in discontinuous trajectories, precluding the calculation of equilibrium time correlation functions for times longer than the exchange time $\delta t_\mathrm{REMD}$.  To improve the sampling  efficiency of REMD, the shortest possible $\delta t_\mathrm{REMD}$ should be used \cite{Levy:PNAS:2007}.  With $\delta t_\mathrm{REMD}$ much shorter than the time scales of slow conformational changes, the rates of conformational changes appear inaccessible to REMD simulations.  To overcome this problem, at least for the special case of a two-state system, an indirect method has recently been proposed in which the two rate coefficients describing the assumed folding/unfolding dynamics are assumed to obey an Arrhenius temperature dependence \cite{Spoel:PRL:2006}.  However, the protein-folding rate often exhibits non-Arrhenius temperature dependence \cite{Oliveberg:PNAS:1995}, and folding intermediates are common. To avoid the resulting problems, master-equation approaches have been described by Levy and co-workers \cite{Andrec:PNAS:2005} in a qualitative, yet insightful analysis.  As a quantitative alternative, REMD has recently been used to estimate the local drift and diffusion coefficients \cite{Yang:JMB:2007} within the framework of coarse diffusion equations \cite{HummerKevrekidis:JCP:2003,hummer2005pdd,Best:PRL:2006}.

Here we show how one can efficiently extract accurate transition rates from REMD simulations, both locally between microscopic conformational states and globally between folded and unfolded conformations (and possible intermediates), without the assumption of a certain temperature dependence of the underlying kinetics. In fact, our method can be used to investigate the Arrhenius or non-Arrhenius character of a particular system. We determine short-time propagators in conformation space to overcome the problems arising from the intrinsically discontinuous character of REMD trajectories \cite{Sriraman:JPCB:2005,Buchete:JPCB:2007}.

We first realize that REMD permits the accurate (and formally exact) calculation of short-time correlation functions.  The initial configurations after a replica exchange (with appropriate velocity assignment) constitute valid representatives of the equilibrium phase-space distributions at the respective temperatures. From the subsequent Hamiltonian dynamics until the next exchange, we can obtain exact correlation functions.  The maximum time scale will be a few $\delta t_\mathrm{REMD}$, given by the longest time between accepted replica exchanges.

Specifically, we here determine the frequency of transitions between conformational states.  From the observed molecular transitions, we construct a master equation describing the dynamics in a conformation space divided into $N$ distinct states.  We later verify that the dynamics in the resulting projected space is captured by a master equation,  $dP_i(t)/dt =  \sum_{j=1}^N \left[ k_{ij} P_j(t) -  k_{ji} P_i(t) \right]$, where $P_i(t)$ is the population in state $i$, and $k_{ij}\geq 0$ is the transition rate from $j$ to $i\neq j$.  In vector-matrix notation, we have $d{\bm{P}}(t)/dt = \bm{K} \bm{P}(t)$, where the $N\times N$ rate matrix $\bm{K}$ has off-diagonal elements $k_{ij}$ and diagonal elements $k_{ii} = - \sum_{j\neq i} k_{ji} < 0$.  The propagators, defined as the probability of being in state $j$ at time $t$ given that the system was in state $i$ at time 0, can be written in terms of the matrix exponential, $p(j,t|i,0) = \left[\exp({\bm{K}t}) \right]_{ji}$.  To estimate the elements of the rate matrix $\bm{K}$ from either long equilibrium simulations or REMD, we use a maximum-likelihood procedure.  We first determine the number $N_{ji}$ of transitions from state $i$ to state $j$ within a time interval $\Delta t$, irrespective of intermediate states.  The log-likelihood of observing transition numbers $N_{ji}$ is \cite{Sriraman:JPCB:2005,Buchete:JPCB:2007}
\begin{eqnarray}
  \ln \mathcal{L} = \sum_{i=1}^{N} \sum_{j=1}^{N}  N_{ji} \ln  p(j,\Delta t | i,0)~.
\end{eqnarray}
To obtain the rate coefficients of the master equation (with upper and lower diagonal elements related by detailed balance), we maximize $\ln\mathcal{L}$ with respect to the $k_{ij}$ \cite{Sriraman:JPCB:2005,Buchete:JPCB:2007}.

Effects of non-Markovian dynamics not captured by the master equation result in a dependence of the rate matrix on the time interval $\Delta t$.  Ultimately, for long lag times $\Delta t$, fast non-Markovian dynamics is effectively suppressed and the propagators are dominated by the slow transitions \cite{Sriraman:JPCB:2005,Best:PRL:2006,Buchete:JPCB:2007}.  However, if $\Delta t$ is short, fast motions lead to improper assignments of conformational states.  As a consequence, the extracted rate matrices tend to predict overly fast conformational relaxation.

\begin{figure}[tb]
\includegraphics{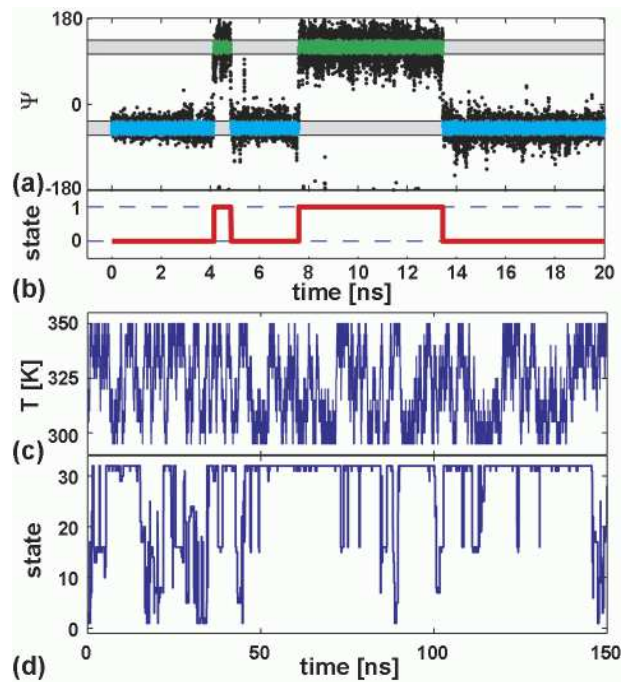}
\caption{\label{fig:01}
REMD simulations. (a) Schematic of transition-path based assignment of conformational states, shown for illustrative purposes in 1D (with the actual assignment done using both $\Phi$ and $\Psi$ \cite{Buchete:JPCB:2007}).   The backbone dihedral angle $\Psi$ of alanine exhibits transitions between helical (blue, $\Psi<0$) and non-helical states (green, $\Psi>0$).  Conformations within narrow regions around the two free energy minima (gray) can be assigned as helical or coil with high confidence. For other conformations (black dots), the assigned state changes only if the trajectory crosses between the well-defined regions, but not on equilibrium excursions that revert without actual crossing. (b) State assignment corresponding to (a). (c) Temperatures sampled by a typical Ala5 replica during a 150 ns REMD simulation. (d) Conformational states sampled by the Ala5 replica during the same run. 
}
\end{figure}

The problem of fast non-Markovian dynamics can be suppressed by assigning the states with the help of transition paths that connect well-defined regions within two conformational cells (Fig.~\ref{fig:01}a-b).  A new state is assigned only if the trajectory crosses from one well-defined region to another.   Fast equilibrium fluctuations in the projected space thus do not lead to a state change.  We showed previously that for peptide folding in long standard molecular dynamics (MD) simulations, this procedure gives accurate rate matrices for observation times $\Delta t$ as short as 1 ps \cite{Buchete:JPCB:2007}.

%%%%%%%%%%%%%%%%%%%%%%%%%%%%%%%%%%
%%%%%%%%%%%%%%%%%%%%%%%%%%%%%%%%%

Here, we adapt this state-assignment procedure to REMD. In a first step, we follow each replica irrespective of exchanges, and identify transition paths for these continuous trajectories to assign states.  In a second step, transition numbers $N_{ji}$ for each of the REMD temperatures are determined from the respective short trajectory segments uninterrupted by replica exchange.  From the $N_{ji}$, we then estimate the coefficients of the master equation through likelihood maximization.

In the following, we demonstrate the general procedure to calculate slow rates from REMD with fast exchange. Master-equation approaches have been used extensively in peptide folding studies \cite{Schutte:JCompPhys:1999,Swope:JPC:2004a,Andrec:PNAS:2005,deGroot:JMB:2001,Becker:JCP:1997,Sriraman:JPCB:2005,buchete2001mfp}. We used the GROMACS 3.3 package \cite{GROMACS3.0} to run both standard MD and REMD simulations for the folding of a short helical peptide, blocked Ala$_5$ (i.e., CH3CO-Ala5-NHCH3), in explicit water \cite{Hummer:PRL:2000,margulis2002hua}.  We used the AMBER-GSS force field \cite{nymeyer2003sfe} ported to GROMACS \cite{sorin2005ehc}, with peptide ($\Phi,\Psi$) torsional potentials modified to reproduce experimental helix-coil equilibria \cite{GarciaSanbonmatsu:PNAS}.  Simulation details can be found in Ref.~\onlinecite{Buchete:JPCB:2007}.

Four independent MD and REMD runs were initiated from different configurations ($11111$ - `all helix', $00000$ - `all coil', $01010$, and $10101$, where $1$ denotes a residue in the helical region of the Ramachandran map, ordered left to right from N to C terminus \cite{Buchete:JPCB:2007}).  The reference MD runs covered 250 ns at two different temperatures (300 and 350 K), for a total combined simulation time of 2 $\mu s$. The 150-ns REMD simulations used 12 replicas spanning the 295-350 K temperature range for a total combined simulation time of 600 ns per replica. Coordinates were saved every 1 ps and REMD exchanges were attempted every $\delta t_\mathrm{REMD} = 5$ ps.  Figure~\ref{fig:01}c shows that the resulting REMD trajectories pass through the whole range of temperatures multiple times.  Each individual trajectory also has a high likelihood to visit most, if not all, of the 32 coarse-grained conformational states (Fig.~\ref{fig:01}d; with 00000 and 11111 corresponding to states 1 and 32 in binary notation plus 1).  In the resulting master equation model, the transition rates $k_{ij}$ are different from zero only if states $i$ and $j$ in binary notation differ by at most one bit, producing the connectivity of a five-dimensional hypercube.

\begin{figure}[tb]
\includegraphics{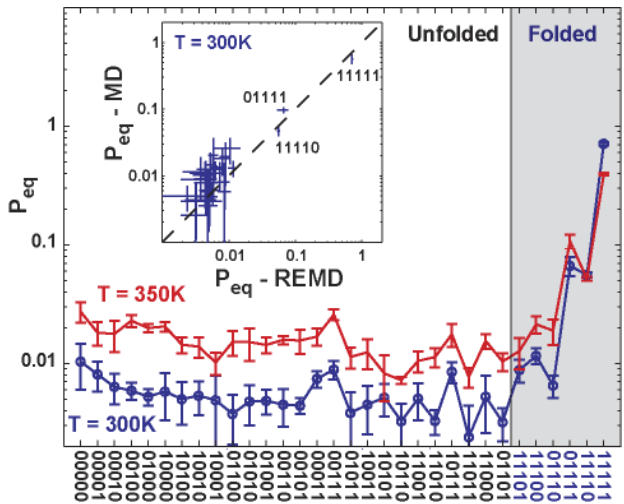}
\caption{\label{fig:02} 
REMD equilibrium populations $P_{eq}$ at 300 and 350 K.  Shading indicates the folded basin. (Inset) Scatter plot of $P_{eq}$ from standard MD and REMD at 300 K.  Error bars indicate standard deviations of the mean. 
}
\end{figure}

Figure~\ref{fig:02} shows the equilibrium populations in each of the 32 conformational states at 300 and 350 K from REMD trajectories.  The  inset illustrates the excellent agreement between equilibrium distributions from MD and REMD at 300 K.  At 350 K, the sampling is more efficient and the agreement even better (data not shown).

%%%%%%%%%%%%%%%%%%%%%%%%%%%%%%%%% 
%%%%%%%%%%%%%%%%%%%%%%%%%%%%%%%%% 

\begin{figure}[tb]
\includegraphics{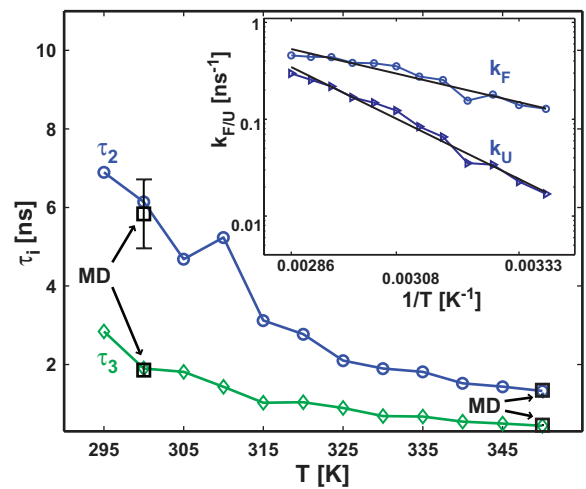}
\caption{\label{fig:03}
Relaxation times $\tau_2$ (circles, blue) and $\tau_3$ (diamonds, green) as a function of temperature.  Open squares show $\tau_2$ and $\tau_3$ from standard MD at 300 and 350 K \cite{Buchete:JPCB:2007}.  (Inset) Folding ($k_F$) and unfolding ($k_U$) rate constants as a function of $1/T$.
}
\end{figure}

\begin{figure}[tb]
\includegraphics{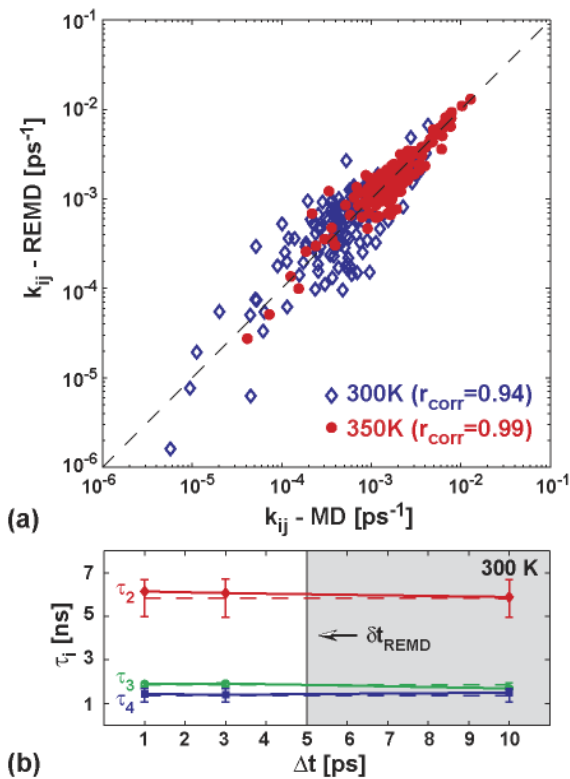}
\caption{\label{fig:04}
  Validation of transition rates estimated from REMD trajectories.  (a) Rates $k_{ij}$ from REMD versus those from standard MD  \cite{Buchete:JPCB:2007} at 300 K (blue) and 350 K (red).  (b) Dependence of the relaxation times $\tau_2$ (top, red), $\tau_3$ (middle, green), and  $\tau_4$ (bottom, blue) on the lag time $\Delta t$ at 300 K. REMD results are shown as symbols connected by solid lines.  Reference values from standard MD are shown as dashed lines with error bars.  Results for $\Delta t > \delta t_\mathrm{REMD}$ were obtained from continuous trajectory segments in which replica exchange attempts had been rejected.}
\end{figure}

Figure~\ref{fig:03} demonstrates that the master equation accurately captures the kinetics.  Shown are the two slowest relaxation times, $\tau_2$ and $\tau_3$, at the 12 temperatures sampled in the REMD runs (where $\tau_i=-1/\lambda_i$, with $\lambda_i$ the ordered eigenvalues of $\bm{K}$). The REMD relaxation times agree perfectly with those obtained from standard MD runs at 300 and 350 K \cite{Buchete:JPCB:2007}.  This agreement holds also for all relaxation times $\tau_i$ (not shown for $i\geq 4$), and the individual coefficients $k_{ij}$ of the master equation, as shown in Fig.~\ref{fig:04}a (with linear correlation coefficients $\geq 0.94$).

%%%%%%%%%%%%%%%%%%%%%%%%%%%%%%%%%%
%%%%%%%%%%%%%%%%%%%%%%%%%%%%%%%%%

%%%%%%%%%%%%%%%%%%%%%%%%%%%%%%%%%%
%%%%%%%%%%%%%%%%%%%%%%%%%%%%%%%%%

From the slowest relaxation time $\tau_2$, and the relative populations in the folded (helical) and unfolded (coil) state of the peptide, we estimate folding and unfolding rates as a function of temperature under the assumption of a two-state relaxation (Fig.~\ref{fig:03} inset).  The 32 states $i$ were assigned as folded or unfolded based on the left-hand eigenvector of $\bm{K}$ corresponding to eigenvalue $\lambda_2$ \cite{Buchete:JPCB:2007,berezhkovskii2004ets} (see Fig.~\ref{fig:02}).  The resulting folded basin consists of all structures with at least one $\alpha$-helical (i,i+4) backbone hydrogen bond among the four N-terminal residues.  Consistent with the assumptions of Ref.~\onlinecite{Spoel:PRL:2006}, we find that the resulting folding and unfolding rates exhibit Arrhenius-like dependence on temperature. The activation energies for folding and unfolding 
are $E_a^F \approx 22.1$ kJ/mol and $E_a^U \approx 46.5$ kJ/mol.

A possible concern is the influence of fast non-Markovian dynamics not taken into account by the master equation model.  We can explicitly probe for such effects by plotting the calculated relaxation times $\tau_i$ as a function of the lag time $\Delta t$ used to determine the propagators.  Figure~\ref{fig:04} shows that the relaxation times from REMD are independent of $\Delta t$ from 1 to 10 ps ($2 \delta t_\mathrm{REMD})$, and agree with the results from standard MD.

We showed how accurate rates for the conformational dynamics of a molecular system can be extracted from REMD simulations.  For a short helical peptide in water, the REMD kinetics was in perfect agreement with that from standard MD. The key elements of the procedure are (1) the suppression of non-Markovian noise by using transition paths in the assignment of states, (2) the calculation of transition numbers $N_{ij}$ on the time scale of replica exchanges, and (3) the construction of a master equation from the $N_{ij}$ using a maximum likelihood procedure.  The formalism is general, and can be adapted to Hamiltonian REMD \cite{Fukunishi:JCP:2002}, resolution exchange \cite{Lyman:PRL:2006}, non-Boltzmann reservoirs \cite{Roitberg:JPCB:2007}, serial replica exchange \cite{Hagen:Berne:JPCB:2007}, etc.

In practical applications, such as protein folding, the combinatorial explosion in the number of states poses a major challenge for large systems. 
%\color{Red}
To reduce the dimension of the master equation, states could be defined by using conformational clustering \cite{chodera2007adm}, subsets of the dihedral-angle coordinates (that produce the most Markovian dynamics), or alternative coordinates such as native or non-native amino-acid contacts or contact fractions, the radius of gyration, or distances between key residues, with our formalism applicable to both discrete and continuous variables \cite{hummer2005pdd}. 
%\color{Black}  
In addition, hierarchical coarse graining \cite{Noe:Smith:JCP:2007} can be used to combine fine and coarse-grained master equations \cite{HummerKevrekidis:JCP:2003,Sriraman:JPCB:2005}.  As a second challenge, the need to collect sufficient transitions at all temperatures to construct a connected master equation could be overcome by assuming that the individual rates $k_{ij}$, but not necessarily the slow relaxations $\tau_i$, satisfy an Arrhenius law. In that way, transitions observed at higher temperatures can be used to estimate the relaxation time scales at lower temperatures, augmented by the accurate equilibrium populations of REMD through the requirement of detailed balance.  Such a procedure is easily implemented within our likelihood-maximization framework by replacing the individual rates with temperature-independent prefactors and activation energies.

\acknowledgments 
%\smallskip
%\skip
%\tiny{
%\scriptsize{
%\footnotesize{
We thank Drs. A. Szabo, A. M. Berezhkovskii, E. Rosta, and R. B. Best for many helpful and stimulating discussions.  This research used the Biowulf Linux cluster at the NIH, and was supported by the Intramural Research Program of the NIDDK, NIH.
%}
%\small{
%We thank A. Szabo, A. M. Berezhkovskii, E. Rosta, and R. B. Best for many helpful discussions. This work 
%used the Biowulf cluster, and 
%was supported by the NIH Intramural Research.
%}

%\bibliography{remd_v28}

\end{document}